\documentclass[submission,copyright,creativecommons]{eptcs}
\providecommand{\event}{Formal Engineering approaches to Software Components and Architectures 2015}
\def\authorrunning{V. Koutsoumpas}
\def\titlerunning{A Formal Approach based on Fuzzy Logic for the Specification of Interactive Systems}

\usepackage[pdftex]{graphicx}
\usepackage{rotating}
\usepackage{wrapfig}
\usepackage{mdwlist}
\usepackage{paralist}
\graphicspath{{../gfx/}}
\DeclareGraphicsExtensions{.pdf,.png,.eps}
\usepackage[cmex10]{amsmath}
\usepackage{esvect}
\usepackage{amssymb}
\usepackage{calrsfs}
\DeclareMathAlphabet{\mathcal}{OMS}{zplm}{m}{n}
\usepackage{amsthm}
\usepackage{mathrsfs}
\usepackage{fixltx2e}
\usepackage{mathtools}
\usepackage{stmaryrd}
\usepackage{algorithmic}
\usepackage{array}
\usepackage[caption=true, font=footnotesize]{subfig}
\usepackage{fixltx2e}
\usepackage{stfloats}
\usepackage{cite}
\usepackage{url}
\usepackage{color}
\usepackage{multirow}

\usepackage{listings}
\usepackage{hyperref}
\usepackage{xcolor}
\usepackage[latin1]{inputenc}
\usepackage{tikz}
\usetikzlibrary{positioning,shapes,arrows,calc,backgrounds,matrix}
\theoremstyle{definition}\newtheorem{definition}{Definition}
\theoremstyle{remark}\newtheorem{example}{Example}
\theoremstyle{plain}\newtheorem{theorem}{Theorem}

\newcommand\HUGE{\@setfontsize\Huge{38}{47}} 

\newcommand\defeq{\stackrel{def}{=}}

\makeatletter
\newcommand{\oset}[3][0.5ex]{%
  \mathrel{\mathop{#3}\limits^{
    \vbox to#1{\kern-0\ex@
    \hbox{$\scriptstyle#2$}\vss}}}}
\makeatother

\begin{document}
\tikzstyle{block} = [draw,  rectangle, 
    minimum height=3em, minimum width=6em]
\tikzstyle{blockRounded} = [draw,  rectangle, 
    minimum height=3em, text width=10em, text ragged,rounded corners, very thick,text centered] 
\tikzstyle{block3} = [draw,  rectangle, 
    minimum height=3em, text width=5em, text ragged,rounded corners, very thick,text centered]  
\tikzstyle{oval} = [draw,  ellipse, 
    minimum height=5em, text width=2.25em,  very thick,text centered ] 
\tikzstyle{decision} = [diamond, draw, fill=white, text width=4em, text badly centered, node distance=1.5cm, inner sep=0pt]
\tikzstyle{blockm} = [draw,  rectangle, 
    minimum height=3em, text width=7em, text ragged, text centered, fill=white,node distance=1.5cm] 
\tikzset{snake it/.style={decorate, decoration={snake, amplitude=0.5mm}}}
\definecolor{myBlue}{RGB}{0,112,192}
\definecolor{myGreen}{RGB}{0,176,80}

\tikzstyle{sum} = [draw,  circle, node distance=1cm]
\tikzstyle{temp} = [coordinate]
\tikzstyle{input1} = [coordinate]
\tikzstyle{input2} = [coordinate]
\tikzstyle{output1} = [coordinate]
\tikzstyle{output2} = [coordinate]
\tikzstyle{pinstyle} = [pin edge={to-,thin,black}]

\title{A Formal Approach based on Fuzzy Logic for the Specification of Component-Based Interactive Systems}
\author{Vasileios Koutsoumpas
\institute{Technische Universit\"at M\"unchen, Munich, Germany}
\email{koutsoum@in.tum.de}}
\maketitle
\begin{abstract}
Formal methods are widely recognized as a powerful engineering method for the specification, simulation, development, and verification of distributed interactive systems. However, most formal methods rely on a two-valued logic, and are therefore limited to the axioms of that logic: a specification is valid or invalid, component behavior is realizable or not, safety properties hold or are violated, systems are available or unavailable. Especially when the problem domain entails uncertainty, impreciseness, and vagueness, the appliance of such methods becomes a challenging task. In order to overcome the limitations resulting from the strict modus operandi of formal methods, the main objective of this work is to relax the boolean notion of formal specifications by using fuzzy logic. The present approach is based on Focus theory, a model-based and strictly formal method for component-based interactive systems. The contribution of this work is twofold: 
\begin{inparaenum}[\itshape i\upshape)]
\item we introduce a specification technique based on fuzzy logic which can be used on top of Focus to develop formal specifications in a qualitative fashion;
\item we partially extend Focus theory to a fuzzy one which allows the specification of fuzzy components and fuzzy interactions. 
\end{inparaenum}
While the former provides a methodology for approximating I/O behaviors under imprecision, the latter enables to capture a more quantitative view of specification properties such as realizability.
\looseness=-1
\end{abstract}
\section{Introduction}
\label{sec:Introduction}
Formal methods are widely recognized as a powerful engineering method for the specification of interactive systems \cite{Broy2001}. They follow the principle of ``\textit{correctness by construction}" and are therefore well suited for security-critical systems \cite{hall2002correctness}. Although the promises of formal methods are well known \cite{Luqi1997FormalMethods}, there are many limitations preventing the usage in industrial software development. The following limitations are generally identified in literature \cite{Sommerville2006,Cerny2014InterfaceSimulationDistances} as the main blockers:
\begin{inparaenum}[\itshape 1\upshape)]
\item \textit{Limited scope}: Formal methods are not well suited to specifying user and  environment interfaces and interactions;
\item \textit{Limited scalability}: As systems increase in size, the time and effort required to develop a formal specification grows disproportionately; \item \textit{Limited expressiveness}: standard formal methods are not capable to quantify values between the ``absolute truth" and the ``absolute false".
\end{inparaenum}

\begin{figure}
\vspace{0cm}
\centering
\includegraphics[scale=0.75]{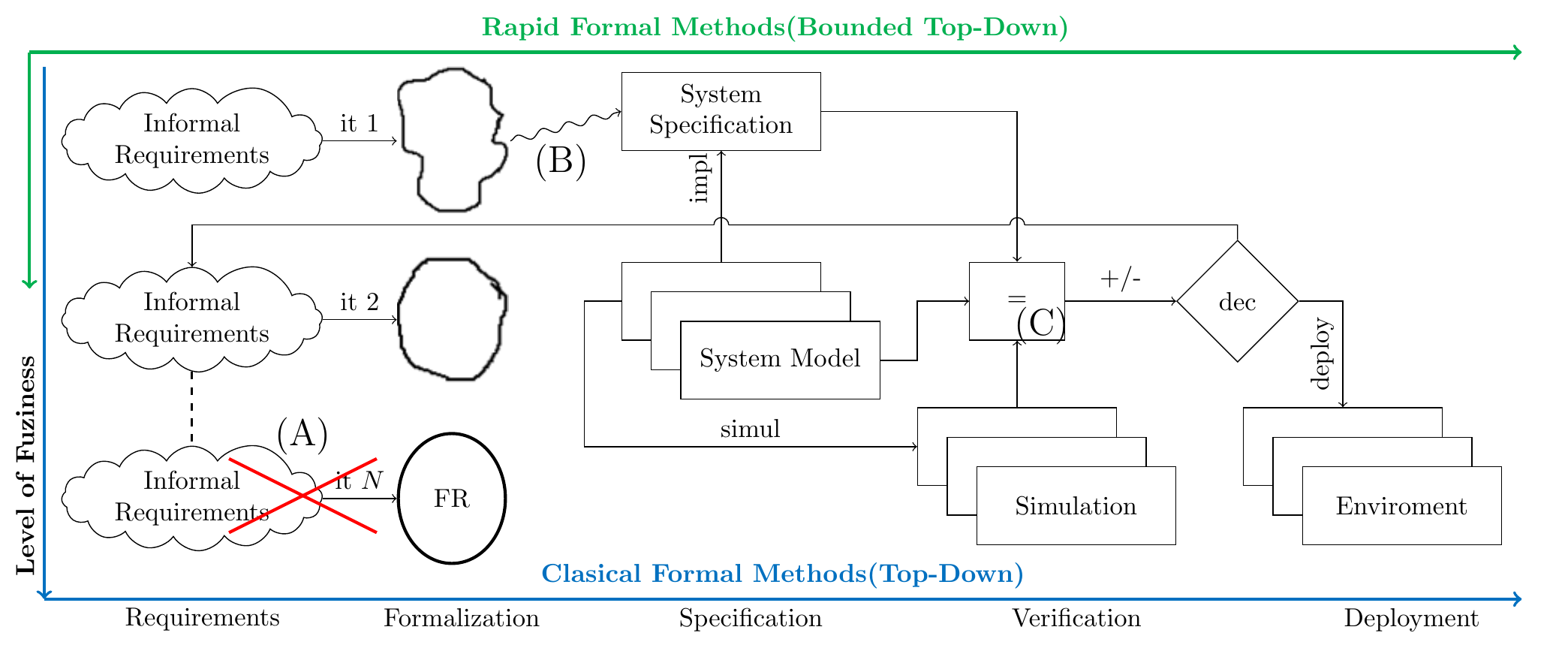}
\caption{(A) Instead of an iterative refinement we suggest to proceed with formal modeling as early as possible (B) We propose a  specification technique for: 1) formalizing qualitative properties of components and; 2) for approximating component behavior in terms of a rule base (C) we provide an equivalence model which allows to capture distances between specifications and systems }
\label{fig:motivation}
\end{figure}

Through the longtime experience obtained within the research projects SPES \cite{Pohl2012SPES}  and E-Energy\footnote{\url{http://www.e-energy.de/en/}}, we empirically confirmed the presence and challenges of the above stated limitations for the avionic, automotive, and smart grid domain. Driven from the individual problems recognized in each domain, there is a natural question whether it is possible to extend standard formal methods to allow on the one hand to speed up the development of specifications while on the other hand the specification should remain formal enough to allow the promises of formal methods such as verification, model checking, etc. To advance this overarching question we distinguish between two major problem categories:  
\paragraph{Problem Statement 1:} Formal methods, such as Focus \cite{Broy2001} or Z \cite{Jacky:1996:WZP:249512}, permit the precise and unambiguous modeling of interactive component behavior. To achieve that, it's necessary to formalize the informal system requirements. Since vagueness, imprecision, and ambiguity are inherent in natural language, the informal system requirements suffer also from this. Thus, a tight feedback loop between detailed requirements specification and formal specification is observed and repeated until the formal specification becomes precise enough to continue with the implementation. Nevertheless, some system problems, particularly those drawn from the systems engineering domain, where the system's context includes user and environment interactions, may be difficult to model in crisp or precise terms. Furthermore, in order to meet the project's time constraints, it may be desirable that formal methods  should commence as early as possible, even though the understanding of parts of the problem domain is only approximate. Hence, the first problem we deal with in this paper is visualized in Figure \ref{fig:motivation} and addresses the research question: How to soften the aforementioned tight feedback loop?
\vspace{-0.5cm}
\paragraph{Problem Statement 2:}Once a formal system specification is defined, standard verification systems (e.g. Isabelle \cite{Nipkow-Paulson-Wenzel:2002}) return a boolean answer that indicates whether a system behavior conforms to its specification. Hence, two distinct behavior clusters are formed, namely that of correct and that of incorrect behaviors. However, not all correct behaviors are equally good, and not all incorrect behaviors are equally bad. Thus, a second research question rises whether it is possible to relax the strict boolean notion of formal methods to capture a more fine grained view as depicted in Figure \ref{fig:motivation} between specification and possible implementations. Such a view, allows for quantitative reasoning about specification properties such as realizability, safety, and liveness, to name only a few.   

\vspace{-0.5cm}
\paragraph{Motivation and Research Objective:}
The identified problems are closely related to the strict mathematical concepts used in formal methods. Most of them are based on crisp sets and on a two-valued logic, and are therefore limited to the axioms of that logic. Many researcher \cite{NeubeckPhilipp2012,Kwiatkowska2004} have successfully applied probabilistic and stochastic approaches to deal with uncertainty resulting from the lack of information. However, there is also another source of uncertainty, resulting from the inability to characterize information. The latter is also the kind of uncertainty we address in this work. In recent years, there is a number of research attempts \cite{Cerny2014InterfaceSimulationDistances,Henzinger2010SimulationDistances,Matthews2000FormalSpec,Matthews2002FuzzyConceptsAndFormalMethodesSample,Henzinger2014ModelMeasuringHybridSystems}, which point out the need for emerging ideas and concepts to overcome these limitations. Indeed, most existing approaches, especially those addressing the second problem are based on distance specification \cite{Henzinger2010SimulationDistances,Henzinger2014ModelMeasuringHybridSystems,Cerny2014InterfaceSimulationDistances}. Their attempt is to relax the boolean notion by defining custom distances for each specification property and to measure the corresponding deviation. The alternative we suggest in this work is an innovative approach where we use fuzzy logic to tackle with this problem. The overall idea is schematically depicted in Figure \ref{fig:motivation} and can be understood as a combination of rapid prototyping with formal methods. We call this engineering method Rapid Formal Methods (RFM).     
The research objective is to establish the basic  foundations and concepts needed towards a complete theory for the specification of fuzzy interactive systems. Such a theory should provide the necessary concepts for developing softer specifications but also for modeling fuzzy interactions. The presentation of a complete theory within this paper is not possible and thus we concentrate on component behavior.\looseness=-1
\vspace{-0.5cm}
\paragraph{Structure:} Section \ref{sec:Foundations} presents the related foundations of Focus and fuzzy set theory. In addition the conventions made for this paper are declared. Section \ref{sec:FuzzyLogiconTopOfFocus} describes how fuzzy logic applies on top of Focus to develop specifications based on qualitative properties. In Section \ref{sec:Fuzzy Components} the concept of fuzzy components is introduced and the necessary formalisms are presented. Section \ref{sec:Related Work} lists the related work and establishes a border between this and other approaches. Finally, Section \ref{sec:Conclusion} concludes the present work and describes possible future directions.
\section{Preliminaries}
\label{sec:Foundations}

\paragraph{Focus Theory.}
We base our approach on Focus~\cite{Broy2001}, a model-based and strictly formal software and systems engineering method for distributed interactive systems. 
The method builds on top of High-Order, two-valued, typed Logic (HOL~\cite{Andrews:1986:IML:42772}), which describes systems in terms of their structure (syntactic) and behavior (semantic).
The system structure is determined by a static hierarchy of components, each defining an interface $ I \blacktriangleright O$ through a set of typed input channels $I \in \mathbb{I}$ and typed output channels $O \in \mathbb{O}$.

The central concept of Focus is that of a stream, which is used to represent communication histories. Let M be a given set of messages. A stream $s$ over the set M is a finite ($M^*$) or an infinite ($M^\infty$) sequence of elements from M. Furthermore, the set of timed streams denoted by $M^{\aleph} =_{def} (M^*)^\infty$ represent an infinite history of finite communications over a channel that are carried out in a discrete time frame. The k-th sequence in a timed stream represents the sequence of messages exchanged on the channel in the k-th time interval. 

Further, different components can be connected through I/O channels to describe component interaction through message exchange. Hence, component behavior is determined by a mapping from the set of possible input histories (streams over input channels $\vv{I}$) to the set of possible output histories (streams over output channels $\vv{O}$). Therefore, the semantic interface of a component is denoted by a set-valued function $F:\vv{I}\rightarrow \wp(\vv{O})$. For example, this mapping can be expressed by means of automata including states and transitions with guards over input histories and actions over output histories, but other description techniques such as table specifications~\cite{Davis1988} are supported in principle as well.

\paragraph{Fuzzy Set Theory.} We assume that the reader has a basic knowledge of fuzzy set theory and fuzzy logic. For a detailed description, we refer to \cite{zadeh1965fuzzy, Zadeh1999PossibilityTheory, Kruse1994FoundationOfFuzzySystems}.

A fuzzy set $\mu$ of $X$ is a function from the reference set $X$ to the unit interval, formally $\mu:X\rightarrow [0,1]$. $\mathcal{F}(X)$ denotes the set of all fuzzy sets of $X$. The value $\mu(x)$ is called degree of truth and the function $\mu$ is called membership function.

\begin{wrapfigure}{r}{5cm}
\vspace{-.85cm}
\includegraphics[scale=1]{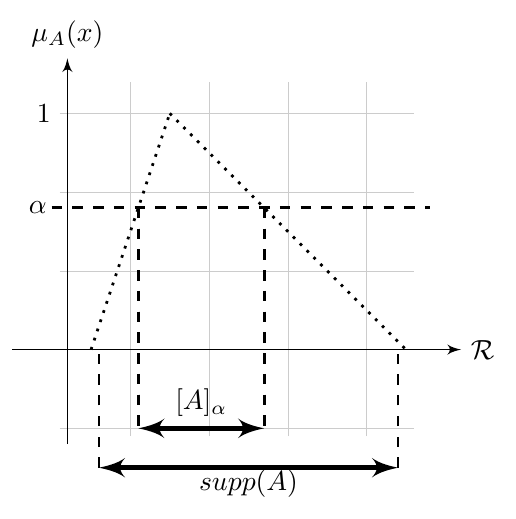}
\vspace{-1cm}
\end{wrapfigure}
\noindent A fuzzy set can be represented by a continuous membership function $\mu$, or by a set $A$ of ordered pairs. The latter is denoted by $A=\{(x,\mu_A(x))\;|\; x\in X\}$. The set $supp(A) =_{def} \{ x\in X \;|\; \mu_A(x)>0 \}$ is called support of A. The set $[A]_{\alpha} =_{def} \{ x\in X \;|\; \mu(x) \geq \alpha \}$ is called $\alpha$-cut of $A$. The fuzzy set $A$ is often denoted by $\{\mu_A(x_1)/x_1,..., \mu_A(x_n)/x_n \}$. Now let $X,Y \subseteq R $ be universal sets, then a fuzzy relation $R$ is a fuzzy set given by $R=\{((x,y),\mu_R(x,y)) \;|\;(x,y)\in X\times Y \}$. Qualitatively, a fuzzy relation can be understood as an expression of the form $R=$ ``x is heavier than y", where $x \in X$, $y \in Y$ and $R \subseteq X \times Y$. Finally, let $R_{1}(x,y) \subseteq X  \times Y$ and $R_{2}(y,z) \subseteq Y \times Z$ be two fuzzy relations. The composition of them is denoted by $R_{1} \circ R_{2}$ defined in $X \times Z$. The membership function of the composed relation is given by the max-min composition denoted by $\mu_{R_{1}\circ R_{2}} = Sup_{y}Min[\mu_{R_{1}(x,y)},\mu_{R_{2}(y,z)}]$.

\paragraph{Conventions.}
Throughout this paper we make usage of some basic operators on streams. Let $s$ be a stream, then $(s.k)$ denotes the $k$-th element of the stream, $s@t$ denotes the element of a timed stream at time point $t$, ($s$$\downarrow$$k$) denotes the sequence of the first k sequences/messages in the stream and $(\#s)$ is the number of elements in s. For an infinite stream $(\#s)=\infty$. Furthermore, we define the functions $\mathit{max}(s)/\mathit{sup}(s)$ and $\mathit{min}(s)/\mathit{inf}(s)$, returning the maximum/supremum and minimum/infimum element of a finite/infinite stream, respectively. By $s_1 \textcircled c s_2$ we denote the concatenation of two streams. In general, messages of any type are supported by streams but for readability we use only the set of real numbers $\mathbb{R}$. Types and sets used in any context, i.e. $x:T$  and $x \in T$, respectively, are by default to be understood as crisp. Fuzzy sets are always stated explicitly. Fuzzy types are recognized by the prefix $\langle \mathcal{F}\_ \rangle$, followed by the type-name. We define the domain and the range of a fuzzy set by $dom.\mu_{A} =_{def} A$ and $rng.\mu_{A} =_{def} \{\mu_{A}(x)\:|\:x\in A\}$. 
\section{Fuzzy Logic on Top of Focus}
\label{sec:FuzzyLogiconTopOfFocus}

\begin{wrapfigure}{r}{5.3cm}
\vspace{-.85cm}
\includegraphics[scale=1.2]{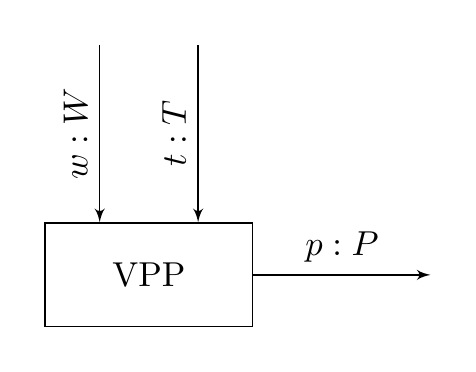}
\vspace{-1.2cm}
\end{wrapfigure}
In this Section we apply fuzzy logic on top of Focus to develop soft specifications for interactive systems. Consider the following simple example of a Virtual Power Plant (VPP) which exchanges weather information with its environment (i.e. weather station) and produces power to supply a network of consumers. A system according to Focus is specified if the syntactic and the semantic interface are fully specified. The former specifies how the system interact $(I \blacktriangleright O)$ with its environment while the latter specifies the behavior of the component denoted by $B:\vv{I}\rightarrow \wp (\vv{O}$). Formalizing the behavior of a component is not always easy. In the given example one first has to decompose the system in its elementary building blocks, for example a set of solar panels. Afterwards the formalization by means of mathematical models like differential equations of each behavior is required. For a detailed  overview on how to apply formal methods to smart grid systems and the coherent challenges, we refer to \cite{Hackenberg2012} and \cite{Hackenberg2014}.\looseness=-1

\subsection{Syntactic Interface - I/O Specification}
First we have to extend the syntactic interface for the introduced example. As illustrated above the $(I \blacktriangleright O)$ of the VPP consists of its input channels $w,t$, its output channel $p$, and the types of messages that are transmitted on them. Messages received on $w/t$ are of type $W/T$ respectively, and messages sent along $p$ are of type P. Since channels are typed, and Focus uses crisp sets to define types, we introduce a new concept namely that of fuzzy properties and fuzzy ports.    

\begin{definition}[Fuzzy Property]
A \textnormal{fuzzy property} $\widetilde{p}$ is a three-tuple $\langle X, \xi, \pi_{\xi} \rangle$, where $X$ is the universe of discourse which can be referenced by $\widetilde{p}$, $\xi$ is a linguistic term which characterizes the property and $\pi_{\xi}:X \rightarrow [0,1] \cup \{ \perp \}$ is the membership function. The value $\pi_{\xi}(x_i)$ is an indicator to what degree the property holds for a given $x_i \in X$. A fuzzy property can be represented by a fuzzy set $X_{\xi}=\{(x,\pi_{\xi}(x))\;|\; x\in X\}$, which is fully specified by the three-tuple. By $\mathcal{P}$ we denote the set of all fuzzy properties.
\end{definition} 

\begin{example}
The tuple $\langle T, HIGH, \pi_{HIGH}\rangle$, where $T=\{t \in \mathbb{R}\;|\; (-30\le t \le 40) \}$ defines a property which describes the high temperature for the VPP. A possible representation could then be  $T_{HIGH}=\{0/15,0.3/20,0.6/25,0.9/30,1/35\}$, where the temperature of $15^{\circ}$C are considered to be high with a degree of truth 0, the temperature of $20^{\circ}$C are considered to be high with a degree of truth 0.3, and so on.\looseness=-1
\label{ex:1}
\end{example}

\begin{definition}[Total Fuzzy Property]
We say that a property $\widetilde{p} = \langle X, \xi, \pi_{\xi} \rangle$ is total, denoted by $Def_{total}(\widetilde{p})$ if: 
\begin{align}
Def_{total}(\widetilde{p}) \Rightarrow  \forall x\in X \;\exists y \in [0,1] : \pi_{\xi}(x)=y
\end{align}
\end{definition}

\begin{definition}[Partial Fuzzy Property]
We say that a property $\widetilde{p} = \langle X, \xi, \pi_{\xi} \rangle$ is partial, denoted by $Def_{partial}(\widetilde{p})$ if: 
\begin{equation}
Def_{partial}(\widetilde{p}) \Rightarrow \exists x \in X : \pi_{\xi}(x)=\perp
\end{equation}
\end{definition}
\noindent In example \ref{ex:1}, the defined property is partial because $\exists t \in T| \pi_{HIGH}(t)=\perp$, e.g. $\pi_{HIGH}(28)=\perp$. Defining total properties is time intensive, mostly because of the partial known interaction with the environment. Additionally, the possible deployment of a system in multiple environments requires to define each property separately for each environment.  We will show later in this paper how to overcome this issues by defining mapping strategies over I/O streams.
 
\begin{definition}[Fuzzy Port]
A fuzzy port ${\Theta}_T$ over a type $T$ is a set of fuzzy properties ${\Theta}_T = \{\widetilde{p}\in \mathcal{P}\}$, which satisfies the following two conditions:
\begin{itemize*}
\item[-]Each property type is a subset of T, formally:
\begin{align}
\forall \widetilde{p} \in {\Theta}_T \rightarrow \widetilde{p}.X \subseteq T \tag{c1}
\end{align}
\item[-] Each property is uniquely characterized by its linguistic term, formally:
\begin{align}
\forall \widetilde{p}_1,\widetilde{p}_2 \in {\Theta}_T \:|\: \widetilde{p}_1\neq \widetilde{p}_2 \rightarrow \widetilde{p}_1.\xi \neq \widetilde{p}_2.\xi \tag{c2}
\end{align}   
\end{itemize*}
\end{definition} 
\noindent A fuzzy port $\Theta_T$ is said to be well defined, only if, c1 and c2 are satisfied, $\Theta_T\vdash c1\land c2$.
Graphically, a fuzzy input/output port is denoted by a white/black circle \( ( \circ ) \)/(\textbullet ), respectively, at the boundary of a component. By $\mathit{IP_S}$/$\mathit{OP_S}$ we denote the set of all fuzzy input/output ports for a given system $S$. Furthermore, by $\widetilde{p}^{\Theta_T}$ we denote the property $\widetilde{p}$ which belongs to the fuzzy port $\Theta_T$. This notation is further generalized also for the elements $\xi^{\Theta_T},\pi_\xi^{\Theta_T}$ of a property. 

Since fuzzy ports are formally specified we can now connect channels with fuzzy ports. I/O channels can be connected to  I/O fuzzy ports respectively through connections. A connection is defined as the binding of a concrete channel to a concrete fuzzy port. Note that not every channel can be connected to a concrete port. This is because ports and channels are specified separately. While the former is a characteristic of the component to be developed the latter may preexist i.e. consider we develop a component for an already existing system. Thus, following connectivity property has to hold:    
\begin{definition}[Connectivity]
A channel $c:C$ can be connected with a fuzzy port $\Theta_T$ only if: $C \subseteq  T$. This property guarantees that each message transmitted over the channel $c$ can be interpreted by the port $\Theta_T$.  
\end{definition}
For the VPP example we define the set of fuzzy input ports $\mathit{IP_{VPP}}=\{\Theta_W,\Theta_T\}$, where $\Theta_W$ = \{$W_{SUNNY}$, $W_{CLOUDY}$\}  (Figure \ref{fig:SyntacticInterfaceVPP}-A) and $\Theta_T$ =\{$T_{LOW}$, $T_{AVERAGE}$, $T_{HIGH}$\}  (Figure \ref{fig:SyntacticInterfaceVPP}-B). The set of fuzzy output ports $\mathit{OP_{VPP}}=\{\Theta_P\}$ contains a single fuzzy port $\Theta_P$ =\{$P_{LOW}$, $P_{AVERAGE}$, $P_{HIGH}$\}  (Figure \ref{fig:SyntacticInterfaceVPP}-C). Figure \ref{fig:SyntacticInterfaceVPP} denotes the defined total properties of each fuzzy port. Intuitively a fuzzy port takes the role of an interpreter. For a given message received at some time point $t$ over a channel $c$, the port gives all possible interpretations for each property. For example, the temperature of {23\textdegree} can be interpreted to be high/average/low with degree of truth 0.2/0.6/0, respectively. Hence, given a port $\Theta_T$ and a measure $t\in T$, a port interpretation defines a total order $\le$ on $\Theta_T$, e.g. $T_{LOW} \le T_{HIGH} \le T_{AVERAGE} |_{t=23}$ 

Concluding, the syntactic interface of a component is fully specified if
\begin{inparaenum}[\itshape 1\upshape)]
\item its I/O channels are specified and additionally to Focus theory
\item  the corresponding fuzzy I/O ports are well defined.\looseness=-1
\end{inparaenum}
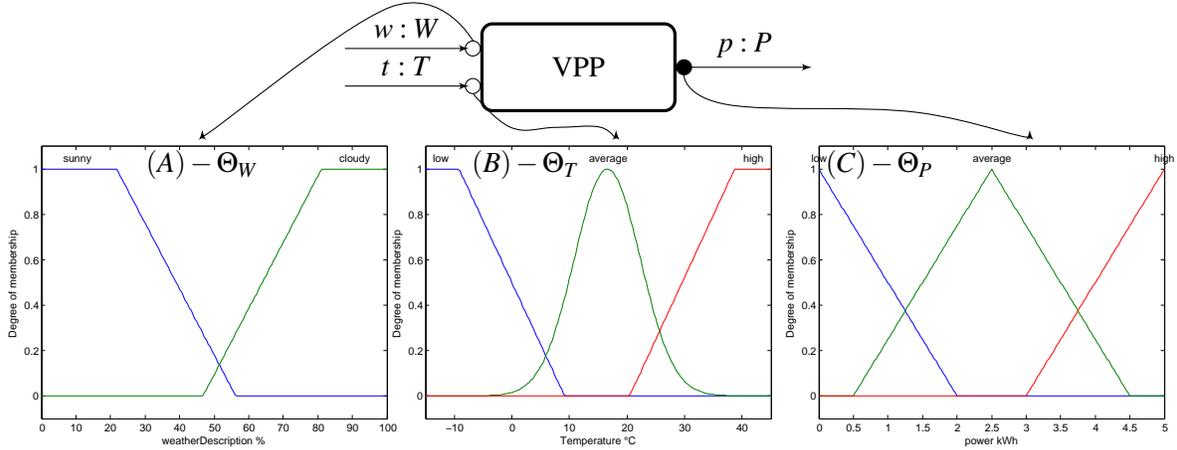
\begin{figure}
\vspace{0cm}
\centering
\begin{tikzpicture}[auto, node distance=5cm,>=latex', scale=1, transform shape]
    
   \node [blockRounded,text width=6em] (vpp) {VPP};

   \draw ($(vpp.west) + (-1mm,+2.5mm)$) circle (1mm);
   \draw ($(vpp.west) + (-1mm,-2.5mm)$) circle (1mm);
   \draw[fill=black] ($(vpp.east) + (1mm,0)$) circle (1mm);
   
   \draw[->] ($(vpp.west) + (-1.8,2.5mm)$) --  node[above] {$w:W$} ($(vpp.west)+ (-1.5mm,2.5mm)$);
   \draw[->] ($(vpp.west) + (-1.8,-2.5mm)$) --  node[above] {$t:T$} ($(vpp.west)+ (-1.5mm,-2.5mm)$);
   \draw[->] ($(vpp.east)$) -- node[above] {$p:P$}  ($(vpp.east) + (1.8,0)$);
   
\node[inner sep=0pt] (whitehead) at (-5,-3)
    {\includegraphics[scale=0.4]{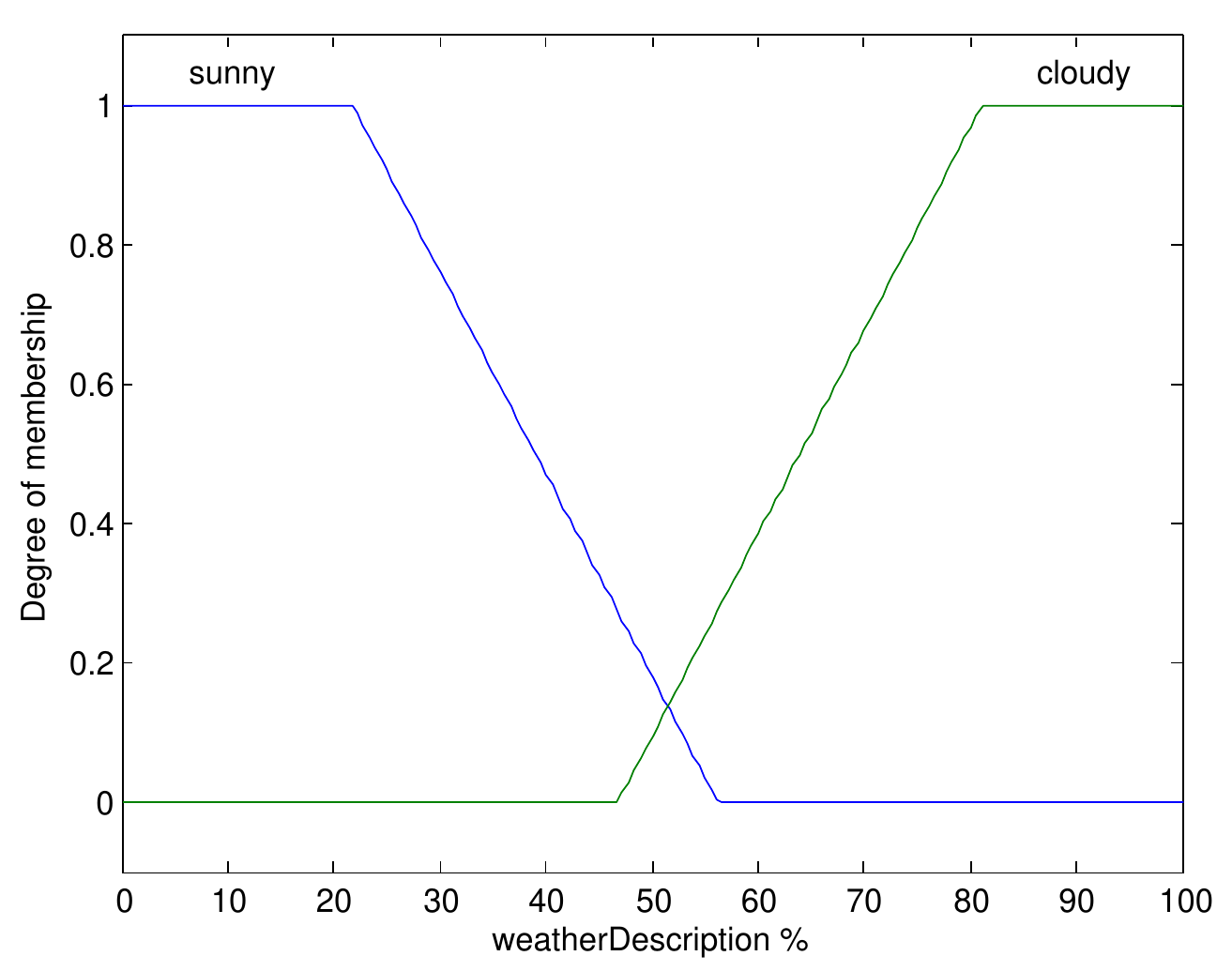}};    
\node[inner sep=0pt] (whitehead) at (0,-3)
    {\includegraphics[scale=0.4]{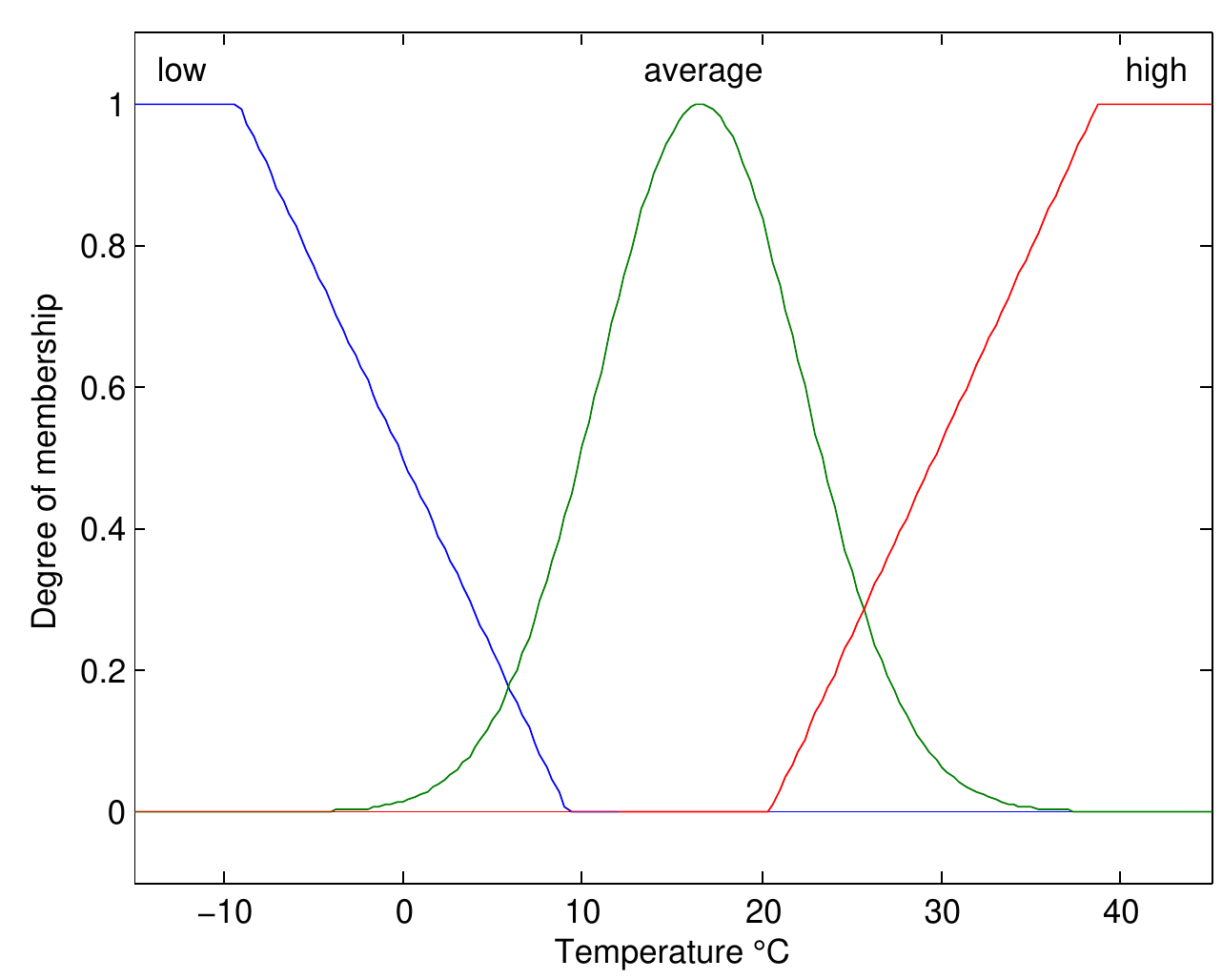}};
\node[inner sep=0pt] (whitehead) at (5.3,-3)
    {\includegraphics[scale=0.4]{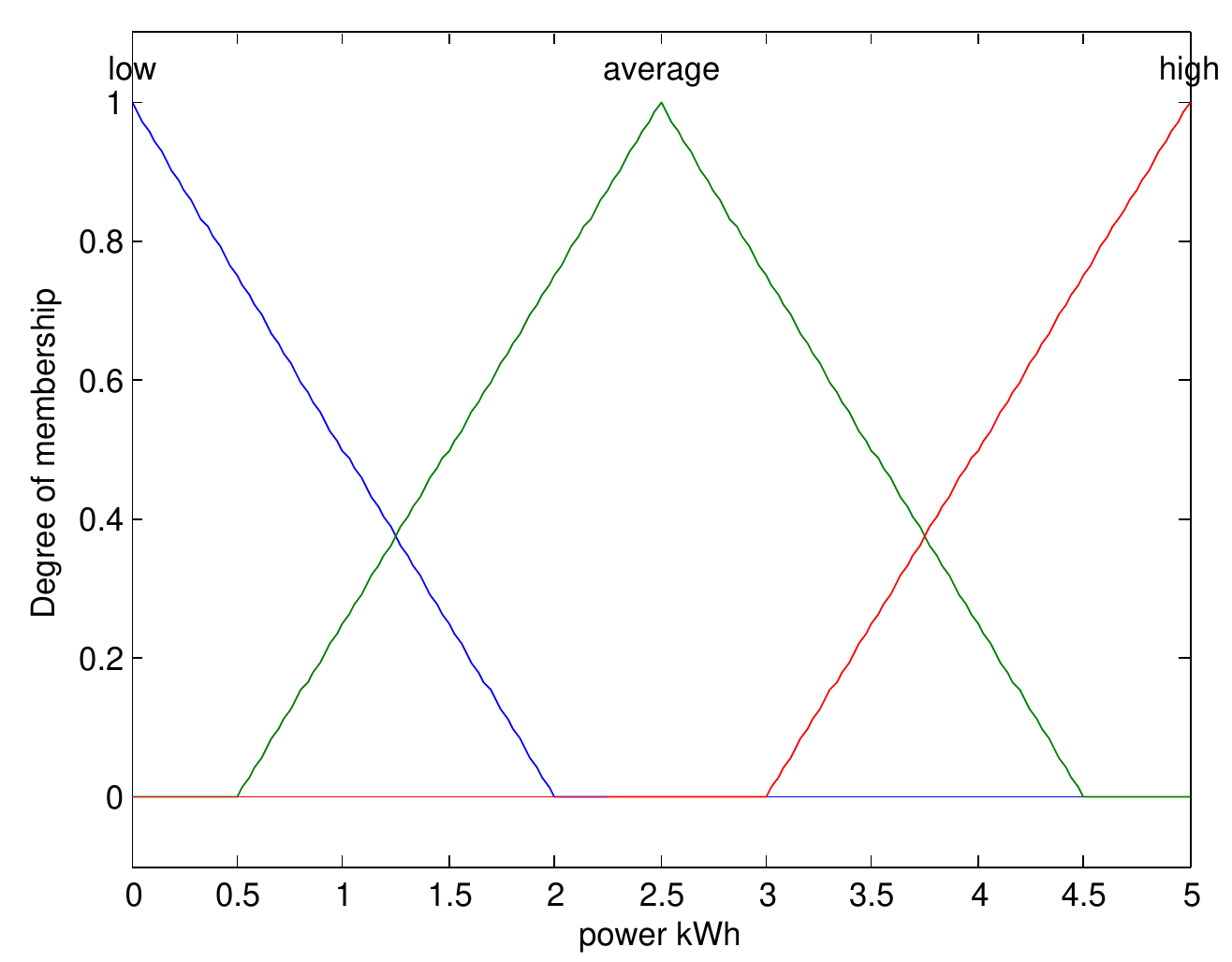}};
\node at (-5,-1.3){$(A)-\Theta_W$};   
\node at (-0.7,-1.3){$(B)-\Theta_T$};   
\node at (4,-1.3){$(C)-\Theta_P$};      
\draw[->]  plot[smooth, tension=.7] coordinates {(1.4,-0.1) (2,-0.5) (4.5,-0.6) (6,-1)};
\draw[->]  plot[smooth, tension=.7] coordinates {(-1.4,-0.35) (-0.9,-0.8) (-0.3,-0.8) (0.3,-0.8) (0.5,-1)};
\draw[->]  plot[smooth, tension=.7] coordinates {(-1.4,0.35) (-2.5,0.8) (-5,-1)};
\end{tikzpicture}
\caption{Syntactic Interface Specification for the VPP}
\label{fig:SyntacticInterfaceVPP}
\vspace{0cm}
\end{figure}
\subsection{Semantic Interface - Behavior Specification} 
\subsubsection{Rule Base Specification.}
After specifying the syntactic interface of a system, we now specify the semantic by a rule base. Let $I=\{i_1:I_1,...,i_n:I_n\}$ and $O=\{o_1:O_1,...,o_m:O_m\}$ be a set of typed I/O channels. Furthermore, let $\mathit{IP}=\{\Theta_{I_1},...,\Theta_{I_n}\}$ and $\mathit{OP}=\{\Theta_{O_1},...,\Theta_{O_m}\}$ represent the well defined fuzzy ports that correspond to the typed I/O channels. For readability, we write $\widetilde{p}^{i}$ instead of  $\widetilde{p}^{\Theta_{I_i}}$ to denote that a property $\widetilde{p}\in \Theta_{I_i}$. Then, a single rule for a specific $o\in O$ has generally the form:\looseness=-1
\begin{align}
\text{                    
$R_{r}^o$: \textbf{if} $i_{1}@t$ is $\xi_{1,r}^{(1)}$ .. \textbf{and} ...\ $i_{n}@t$ is $\xi_{n,r}^{(n)}$ \textbf{then} $o@(t+1)$ \textbf{is} $\xi_{r}$, $r=1,..,k$
}
\label{equ:ruleBase}
\end{align}
\noindent where $\xi_{1,r}^{(1)},...,\xi_{n,r}^{(n)}$, and $\xi_{r}$ represent the linguistic terms that correspond to the fuzzy properties of a fuzzy port such that $\xi_{j,r}^{(j)} = \tilde{p}.\xi \;|\; \tilde{p} \in \Theta_{I_j}, j=1,...,n$ and $\xi_{r} = \tilde{p}.\xi \;|\; \tilde{p} \in \Theta_{O}$.
 
\subsubsection{Behavior Specification} 

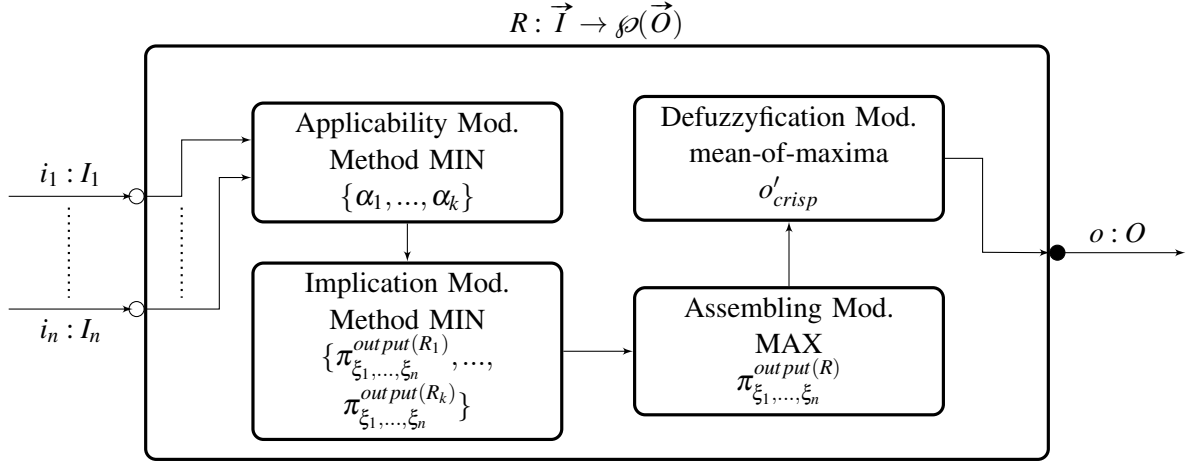
\begin{figure}
\vspace{0mm}
\centering
\begin{tikzpicture}[auto, node distance=5cm,>=latex', scale=1, transform shape]
   \node (system) at (4,2) [draw,very thick,minimum width=12cm,minimum height=5.5cm, rounded corners] {};
   \node(as)  at ($(system.north) + (0,0.3)$) (r) {$R:\vv{I}\rightarrow \wp(\vv{O})$};
   
   \node [blockRounded] at ($(system.west) + (3.5,1.2)$) (c1) {Applicability Mod.\\
   Method MIN\\
   $\{\alpha_{1},...,\alpha_{k} \}$
   };
   
   \node [blockRounded] at ($(c1.south) + (0,-1.7)$) (c2) {Implication Mod.\\
   Method MIN\\
   $\{\pi^{output(R_1)}_{\xi_1,...,\xi_n},...,$\\$ \pi^{output(R_k)}_{\xi_1,...,\xi_n} \}$
   };

   \node [blockRounded] at ($(c2.east) + (3,0)$) (c3) {Assembling Mod.\\
   MAX \\
   $\pi^{output(R)}_{\xi_1,...,\xi_n}$
   };

\node [blockRounded] at ($(c3.north) + (0,1.7)$) (c4) {Defuzzyfication Mod.\\
   mean-of-maxima \\
   $o'_{crisp}$
   };

   \draw ($(system.west) + (-1mm,+7.5mm)$) circle (1mm);
   \draw ($(system.west) + (-1mm,-7.5mm)$) circle (1mm);
   \draw[fill=black] ($(system.east) + (1mm,0)$) circle (1mm);
  
\draw[thick, dotted] ($(system.west) + (-10mm,+6mm)$)  -- ($(system.west) + (-10mm,-6mm)$) ;

\draw[thick, dotted] ($(system.west) + (5mm,+6mm)$)  -- ($(system.west) + (5mm,-6mm)$) ;

   \draw[->] ($(system.west) + (-1.8,7.5mm)$) --  node[above] {$i_1:I_1$} ($(system.west)+ (-1.5mm,7.5mm)$);
   \draw[->] ($(system.west) + (-1.8,-7.5mm)$) --  node[anchor=north] {$i_n:I_n$} ($(system.west)+ (-1.5mm,-7.5mm)$);
   \draw[->] ($(system.east)$) -- node[above] {$o:O$}  ($(system.east) + (1.8,0)$);

\draw[->] ($(system.west) + (0,+7.5mm)$) --  ($(system.west) + (0.5,+7.5mm)$) -- ($(system.west) + (0.5,+1.5)$) -- ($(c1.west) + (0,+3mm)$);

\draw[->] ($(system.west) + (0,-7.5mm)$) --  ($(system.west) + (1,-7.5mm)$) -- ($(system.west) + (1,+1)$) -- ($(c1.west) + (0,-2mm)$);

\draw[->] ($(c1.south) $) -- ($(c2.north) $);
\draw[->] ($(c2.east) $) -- ($(c3.west) $);
\draw[->] ($(c3.north) $) -- ($(c4.south) $);

\draw[->] ($(c4.east) $) -- ($(c4.east) +(0.5,0)$) -- ($(c4.east) +(0.5,-1.25) $) -- ($(system.east) $);

\end{tikzpicture}
\caption{Behavior interpretation of a rule based specification}
\label{fig:SemanticInterfaceGlasBox}
\vspace{0cm}
\end{figure}

In the following we explain how the behavior function can be defined. Figure \ref{fig:SemanticInterfaceGlasBox} depicts the required modules for the behavior specification. A tuple $\langle i_1@t,...,i_n@t \rangle \in I_1 \times ... \times I_{n}$ denotes the measured input picked up by the syntactic interface at some time point $t$. For each rule $R_{r}$ in the rule base we determine the degree to which the measured input fulfills the premise of the rule, called degree of applicability $\alpha_{r}=min\{\pi_{\xi_{1,r}}^{(1)}(i_1),...,\pi_{\xi_{n,r}}^{(n)}(i_n)\}$. The applicability degrees are passed to the implication module where each rule $R_r$ implies for the measured input the fuzzy output set $\pi^{output(R_r)}_{\xi_1,...,\xi_n}: O \rightarrow [0,1]$,
$o \longmapsto min\{\alpha_r,\pi_{\xi_{r}}(o)\}$. The output of a rule $R_r$ is a fuzzy set of output values obtained by cutting of the fuzzy set $\pi_{\xi_r}$ at the level of applicability $\alpha_r$. The results are passed to the assembling module which combines all calculated fuzzy output sets (one for each rule) into a single fuzzy output set by determining the maximum $\pi^{output(R)}_{\xi_1,...,\xi_n}: O' \rightarrow [0,1]$,
$o' \longmapsto \displaystyle\max_{0 \leq r \leq k} \{\pi^{output(R_r)}_{\xi_1,...,\xi_n}\}$. The fuzzy output set is passed to the defuzzyfication module which decides for a crisp value $o'_{crisp}$ by selecting the value with the maximum membership degree. In case where more values have the same degree the mean of maxima is selected. Finally, the crisp output value is passed to the output stream $o$. In case of multiple output channels the above procedure is repeated for each $o_i \in O, i=1,...,m$. Thus, the behavior of a system $S$ is fully specified by the set of all output specifications $R_{S}=\{R^{o_1},...,R^{o_m}\}$. Given a set of timed input streams, the output streams are evaluated according to $R_S$ for each time point.

\begin{example}
For the example depicted in Figure \ref{fig:SyntacticInterfaceVPP}, let the fuzzy properties be defined according to the following scheme:
$T_{HIGH} = \{\frac{0}{10},\frac{0.4}{20},\frac{0.6}{25},\frac{0.8}{30},\frac{1}{35}\}$, $T_{LOW}  =\{\frac{0.2}{20},\frac{0.4}{15},\frac{0.6}{10},\frac{0.8}{5},\frac{1}{0}\}$,
$W_{SUNNY} =\{\frac{0}{80},\frac{0.4}{60},\frac{0.6}{40},\frac{0.8}{20},\frac{1}{0}\}$, $W_{CLOUDY} =\{\frac{0}{20},\frac{0.4}{40},\frac{0.6}{60},\frac{0.8}{80},\frac{1}{100}\}$,
$P_{HIGH} =\{\frac{0}{1},\frac{0.4}{2},\frac{0.6}{3},\frac{0.8}{4},\frac{1}{5}\}$,
$P_{LOW} =\{\frac{0}{4},\frac{0.4}{3},\frac{0.6}{2},\frac{0.8}{1},\frac{1}{0}\}$.

\noindent Furthermore, let $R^{p}$ be the rule base specification containing the following rules: 
\begin{align*}
& \text{$R_1$: \textbf{if} $t$ \textbf{is} $HIGH$ \textbf{and} $w$ \textbf{is} $SUNNY$ \textbf{then} $p$ \textbf{is} $HIGH$}\\
& \text{$R_2$: \textbf{if} $t$ \textbf{is} $LOW$ \textbf{and} $w$ \textbf{is} $CLOUDY$ \textbf{then} $p$ \textbf{is} $LOW$}
\end{align*} 
Given the tuple $\langle t@t_1,w@t_1\rangle = \langle 20,40\rangle$ denoting the measured input at time point $t_1$, we are seeking for the output $p$. As a first step we calculate the degree of applicability for each rule. Thus, $\alpha_{1}=min\{\pi_{HIGH}^{(1)}(20),\pi_{SUNNY}^{(2)}(40)\}= min\{0.4,0.6\}=0.4$ and $\alpha_{2}=min\{\pi_{LOW}^{(1)}(20),\pi_{CLOUDY}^{(2)}(40)\}= min\{0.2,0.4\}=0.2$. By cutting of the fuzzy sets $P_{HIGH}$, $P_{LOW}$ to the degree of applicability $\alpha_{1}$ and $\alpha_{2}$, respectively, we get the output value of each rule: $\pi^{output(R_1)}_{HIGH,SUNNY}= \{\frac{0}{1},\frac{0.4}{2},\frac{0.4}{3},\frac{0.4}{4},\frac{0.4}{5}\}$ and $\pi^{output(R_2)}_{LOW,CLOUDY}= \{\frac{0}{4},\frac{0.2}{3},\frac{0.2}{2},\frac{0.2}{1},\frac{0.2}{0}\}$. By assembling the fuzzy outputs of each rule we get:
$\pi^{output(R)}$=$\{\frac{max(0.2,0)}{0}$, $\frac{max(0.2,0)}{1}$, $\frac{max(0.2,0.4)}{2}$, $\frac{max(0.2,0.4)}{3}$, $\frac{max(0,0.4)}{4}$,  $\frac{max(0,0.4)}{5}\}$ = $\{\frac{0.2}{0}$,$\frac{0.2}{1}$,$\frac{0.4}{2}$,$\frac{0.4}{3}$,$\frac{0.4}{4}$, $\frac{0.4}{5}\}$. Finally, applying the mean of maxima we get $o=3.5$ which is the crisp output that is passed to the output channel $p$.  
\end{example}

\begin{theorem}
Every rule based behavior specification $R_{S}=\{R^{o_1},...,R^{o_m}\}$, where $R^{o_i}$ is of the generally form given by equation \ref{equ:ruleBase}, has a deterministic behavior interpretation $R:\vv{I}\rightarrow \wp(\vv{O})$, which defines a total deterministic Moore machine $(\varDelta,\varLambda)$ with transition function:
\begin{equation}
\varDelta:(\Sigma \times (I\rightarrow M^*)) \rightarrow \wp(\Sigma \times (O \rightarrow M^*))
\label{equ:mooreMachine}
\end{equation}  
\end{theorem}
The above theorem states that despite the fact that the rule based behavior specification relies on fuzzy properties, the component behavior from a black box point of view is not fuzzy at all. This implies, that the abstraction from a rule based behavior specification leads to a crisp deterministic interface behavior $R$. Consequently, tools like Autofocus \cite{AutoFocus} and theorem provers like Isabelle \cite{Nipkow-Paulson-Wenzel:2002} can be further used for behavior analysis.

\subsection{Mapping Strategies}
The definition of total properties requires a total mapping from the reference set to the unit interval. This mapping may be achievable for static properties such as the speed of a car. However, most properties especially when modeling complex systems with environmental interactions are in nature not static. How high temperature should be interpreted depends highly on the geographically location the system will be deployed in. Furthermore, the temperature of $15^{\circ}$C may considered to be high in winter but only average in summer. Therefore, properties can be also time dependent. To deal with location and time dependency of properties we introduce the concept of mapping strategies. Such a strategy defines the membership function of a property according to the observed history of a channel. Thus, the property adapts to the location of a component. Additionally, a threshold for the history length may be declared to consider only recent interactions, this guarantees a smooth adaption of the membership function over time.

\begin{definition}[Mapping Strategy]
A mapping strategy for a given property $\widetilde{p}= \langle X,\xi,\pi_\xi \rangle$(partial or total) is a high order function over a stream to a membership function for that property, formally: 
\end{definition}
\begin{equation}
 mapstr_\xi:Stream \; X, \mathbb{N}\cup \{\infty \}  \rightarrow (\pi_\xi: X \rightarrow [0,1])
\end{equation}

\begin{example}[Mapping Strategy]
For the VPP example the signature of a concrete mapping strategy for the property average temperature $T_{AVERGAGE}$ could be declared as:
\begin{align*}
\mbox{ \textbf{fct} } & mapstr_{T_{AVERAGE}}(t: \mbox{Stream T, \textit{n}: Nat}) \mbox{ \textbf{fct} }\pi_\xi(x:\mbox{T})\{ \\
 & \text{\textbf{ret} gaussmf(min(t$\downarrow$n), max(t$\downarrow$n))}\}
\end{align*}
\end{example}
\section{Fuzzy Components}
\label{sec:Fuzzy Components}
\begin{figure}[h!]
\centering
\vspace{0cm}
\begin{tikzpicture}[auto, node distance=5cm,>=latex', scale=1, transform shape]
    
   \node [block3] (c1) {b};
   \node [block3, right = of c1] (c2) {$\hat{b}$};

   \draw ($(c1.west) + (-1mm,+2.5mm)$) circle (1mm);
   \draw ($(c1.west) + (-1mm,-2.5mm)$) circle (1mm);
   \draw[fill=black] ($(c1.east) + (1mm,0)$) circle (1mm);
   
   \draw[->] ($(c1.west) + (-1.8,2.5mm)$) --  node[above] {$i_1:I_1$} ($(c1.west)+ (-1.5mm,2.5mm)$);
   \draw[->] ($(c1.west) + (-1.8,-2.5mm)$) --  node[anchor=north] {$i_2:I_2$} ($(c1.west)+ (-1.5mm,-2.5mm)$);
   \draw[->] ($(c1.east)$) -- node[above] {$o:O$}  ($(c1.east) + (1.8,0)$);

  \draw ($(c2.west) + (-1mm,+2.5mm)$) circle (1mm);
  \draw ($(c2.west) + (-1mm,-2.5mm)$) circle (1mm);
  \draw[fill=black] ($(c2.east) + (1mm,0)$) circle (1mm);
      
  \draw[->] ($(c2.west) + (-1.8,2.5mm)$) --  node[above] {$i_1:\mathcal{F}I_1$} ($(c2.west)+ (-1.5mm,2.5mm)$);
  \draw[->] ($(c2.west) + (-1.8,-2.5mm)$) --  node[anchor=north] {$i_2:\mathcal{F}I_2$} ($(c2.west)+ (-1.5mm,-2.5mm)$);
  \draw[->] ($(c2.east)$) -- node[above] {$o:\mathcal{F}O$}  ($(c2.east) + (1.8,0)$);
   
  \draw [->, bend angle=25, bend left]  (c1.north) to (c2.north);
  \node at (3.5,1.3){extension}; 
\end{tikzpicture}
\caption{Fuzzy Components and Fuzzy Behavior}
\label{fig:FuzzyComponents}
\vspace{0cm}
\end{figure}
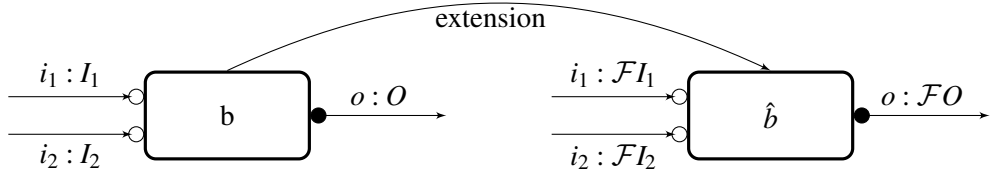

In Section \ref{sec:FuzzyLogiconTopOfFocus} we showed that fuzzy logic is well suited for modeling soft properties and develop rule based specifications. We proved that the abstraction of a rule based behavior specification leads to a crisp deterministic interface behavior $R:\vv{I}\rightarrow \vv{O}$. However, not all correct behaviors are equally good, and not all incorrect behaviors are equally bad. Thus, we introduce the concept of fuzzy components and fuzzy behavior of them. This description yields a quantitative reasoning about component behaviors. Figure \ref{fig:FuzzyComponents} depicts the extension of a component with deterministic behavior $b:\vv{I}\rightarrow \vv{O}$ to a fuzzy component with fuzzy behavior $\hat{b}:\oset{\leadsto}{I}\rightarrow \oset{\leadsto}{O}$ which is the subject of this section.

\subsection{Basic Adaption}
To enable fuzzy component behavior first we have to extend Focus theory in order to deal with fuzzy types. Thus, we introduce the notion of fuzzy types/channels. The prefix symbol $\mathcal{F}$ defines a fuzzy type as a total function from the crisp reference type $T$ to the unit interval [0,1], denoted by $\mathcal{F}T:T \rightarrow [0,1]$. Now, let a set $T_F$ of fuzzy types $\mathcal{F}T$ be given. By $C_F$ we denote the set of fuzzy channels. Furthermore, we assume that we have given a fuzzy type assignment for the fuzzy channels: $f\_type:C_F \rightarrow T_{F}$.  Given a set $C_F$ of fuzzy channels, a valuation or history of a fuzzy channel is denoted by:
\begin{equation}
{\stackrel{\leadsto}{C_F}}=\{x:C_F \rightarrow M^{\mathcal{N}}:\forall c\in C_F: x.c \in \{ dom.(f\_type(c))\}^{\mathcal{N}}   \}
\end{equation}
A valuation of a fuzzy channel $x\in {\stackrel{\leadsto}{C_F}} $ associates a stream $s$ of elements of type $dom.f\_type(c)$ with each fuzzy channel $c \in C_F$. Throughout this chapter we work with a simple notation for streams over fuzzy channels which is described in the following. By $s.j$ we denote the j-th element of the stream $s$ and by $acc_c(s.j)$ we denote the degree of membership of $s.j$ in $dom.f\_type(c)$. Informally, the value $acc_c(s.j)$ tell us to what degree element $s.j$ is accepted by channel $c$. If we combine two elements $(s.j$, $s.k \;|\; j,k \in \mathbb{N} \land j \neq k)$ of a stream, their combination is rated according to the following scheme:
\begin{equation}
\begin{aligned}
& \mbox{ Lower: } acc_{\downarrow}(s.j,s.k)=(s.j \land s.k) = min\{(acc_c(s.j),acc_c(s.k)\}\\
& \mbox{ Upper: } acc_{\uparrow}(s.j,s.k)=(s.j \lor s.k) = max\{(acc_c(s.j),acc_c(s.k)\}
\end{aligned}
\label{equ:bounds}
\end{equation}
For a finite number of elements in a stream we define analogously the acceptance degree of a stream $s$ by:
\begin{equation}
\begin{aligned}
 acc_{\downarrow}(s)= \min_{0 \leq j \leq \#s} \{ acc(s.j) \} \; | \;
 acc_{\uparrow}(s)= \max_{0 \leq j \leq \#s} \{ acc(s.j) \}
\end{aligned}
\end{equation}
Since streams can have an infinite number of elements the above scheme converts to following equations for the infinite case:
\begin{equation}
\begin{aligned}
 acc_{\downarrow}(s)= \inf_{0 \leq j \leq \#s=\infty} \{ acc(s.j) \} \; | \;
 acc_{\uparrow}(s)= \sup_{0 \leq j \leq \#s=\infty} \{ acc(s.j) \}
\end{aligned}
\end{equation}
Furthermore, we can combine not only elements of the same stream but also from different streams as well using the scheme above with following replacement in equation \ref{equ:bounds} $(s.j/s1.j,s.k/s2.k)$. Thus, two or more streams can be combined in order to evaluate the upper and lower acceptance bounds. It is noteworthy to mention that the acceptance degree is not limited to the specified upper and lower bounds in this paper. A statistical representation for the acceptance degree is possible as well (e.g. $\overline{acc(s)}=\frac{1}{\#s}\sum\limits_{j=0}^{\#s}acc(s.j)$). Which representation is best suited depends highly on the system characteristics. Hence, while for a fault tolerant system like a VPP some could prefer the statistical mean representation for a safety critical system like an airplane the lower and upper bounds seems to be more appropriate. Finally, having established a strict notion for fuzzy types, channels and stream processing we introduce the notion of a fuzzy syntactic interface of a component:

\begin{definition}[Fuzzy syntactic interface]
Given a set of fuzzy input channels $I_F$ and a set of of fuzzy output channels $O_F$ we introduce the notion of a fuzzy syntactic interface of a component by $(I_F,O_F)$ or symbolic $(I_F \widetilde{{\blacktriangleright}} O_F)$. 
\end{definition} 

\subsection{Fuzzy Extension}

\begin{theorem}[Fuzzy Type Extension]
Let $f:I^n \rightarrow O$ be a mapping from typed inputs $(i_1:I_1,...,i_n:I_n)$ to a single typed output $o:O$. If the input becomes fuzzy through a fuzzy type assignment of the form $(i_1:\mathcal{F}I_1,...,i_n:\mathcal{F}I_n)$  then the fuzzy type extension of $O$ is given by:
\begin{align}
 \mathcal{F}O(o) \defeq  sup\{& min\{ \mathcal{F}I_1(i_1),...,\mathcal{F}I_n(i_n))\}|\nonumber \\
   & (i_1,...,i_n)\in I^n \mbox{ and } o=f(i_1,...,i_n)\}
\end{align}
\label{theorem:ExtensionPrinciple}
\end{theorem}
\vspace{-0.5cm}
\begin{example}[Stateless Fuzzy Behavior]
We show how the extension principle is applied to a stateless adder with deterministic behavior $o=f(i_1,i_2)$. Let $i_1:I_1$ , $i_2:I_2$ and $o:O$ be of type $I_1=\{2,3,4\}$ , $I_2=\{6,7,8\}$ and $O=\{8,9,10,11,12\}$, respectively. We are seeking for the fuzzy output type $\mathcal{F}O$ if the input of $f$ becomes fuzzy typed.

\begin{minipage}{.45\textwidth}
\begin{flalign*}
 \text{\textbf{fct} }   f=&(i_1: \text{I\textsubscript 1}, i_2: \text{I\textsubscript 2)} \mbox{ \textbf{out} } o:\text{O} \{ && \\
& \text{\textbf{ret} }i_1+i_2;  \}  
\end{flalign*}
\end{minipage}%
\begin{minipage}{.5\textwidth}
\begin{align*}
\text{\textbf{fct} }  \hat{f}=&(i_1: \mathcal{F}I_1, i_2: \mathcal{F}I_2) \mbox{ \textbf{out} } o:\text{\textbf{?}}  \{ \\
& \text{\textbf{ret} }i_1+i_2;  \}  
\end{align*}
\end{minipage}
\vspace{0.2cm}

\noindent Let, $\mathcal{F}I_1$= \{0.5/2, 1/3, 0.5/4\} be a fuzzy type representing the "fuzzy 3"  and  $\mathcal{F}I_2$= \{0.5/6, 1/7, 0.5/8\} another fuzzy type representing the "fuzzy 7". Now, according to Theorem \ref{theorem:ExtensionPrinciple}:
\begin{align*}
\mathcal{F}O(o)=sup\{min\{\mathcal{F}I_1(i_1),\mathcal{F}I_2(i_2))\: |\: i_1\in I_1, i_2 \in I_2 \mbox{ and } o=f(i_1,i_2) \}
\end{align*}
For $i_1+i_2=9$ we receive:
\begin{align*}
\mathcal{F}O(i_1+i_2=9) &= max\{min(\mathcal{F}I_1(3),\mathcal{F}I_2(6)), min(\mathcal{F}I_1(2),\mathcal{F}I_2(7))\} \\
&=max(min(1,0.5),min(0.5,1))=0.5
\end{align*}
Repeating for all $o \in O$ we obtain $ \mathcal{F}O= \{0/8,0.5/9,1/10,0.5/11,0/12 \}$, which is the fuzzy type representing the "fuzzy 10" depicted in figure \ref{fig:fuzzyBehavior}-A.
\label{ex:statelessBehavior}
\end{example}

\subsection{Fuzzy Component Behavior}

Recall from section \ref{sec:FuzzyLogiconTopOfFocus} where component behavior was denoted by $B:\vv{I}\rightarrow \wp (\vv{O}$), meaning that input histories $\vv{I}$ are mapped to all possible output histories $\vv{O}$ over the set-valued function $B$ we turn to the motivation of a general method which enables the mapping of fuzzy input histories to all possible fuzzy output histories over a set-valued function $\hat{B}$, denoted by  
$\hat{B}:\oset{\leadsto}{I} \rightarrow  \wp(\oset{\leadsto}{O})$. 
A fuzzy behavior $\hat{B}$ is called deterministic if $\hat{B}(x)$ is a one element set for each fuzzy input history $x$. Such a behavior is equivalent to a function $\hat{b}:\oset{\leadsto}{I} \rightarrow \oset{\leadsto}{O}$ where $\hat{B}(x)=\{\hat{b}(x)\}$.

\begin{definition}[Fuzzy Behavior Extension]
Let $b:\vv{I}\rightarrow  \vv{O}$ be a mapping from input histories $\vv{I}$ to output histories $\vv{O}$. The fuzzy extension of $b$ is given by:
\begin{align*}
 \hat{b}: \oset{\leadsto}{I} \rightarrow \oset{\leadsto}{O} 
\end{align*}
where $\forall o \in O$ we apply the fuzzy type extension theorem \ref{theorem:ExtensionPrinciple}.
\label{def:BehaviorExtension}
\end{definition}

\begin{definition}[$\alpha$-Realizability]
A fuzzy $I/O$ behavior $\hat{B}$ is called \textnormal{$\alpha$-realizable}, if there exist a total function $\hat{b}: \oset{\leadsto}{I} \rightarrow \oset{\leadsto}{O}$ such that:
\begin{align}
\forall x \in \oset{\leadsto}{I} : \hat{b}(x) \in \hat{B}(x)\land acc(\hat{b}(x))\ge \alpha
\end{align}
$[\hat{b}]_{\alpha}$ is called an $\alpha$-\textnormal{realization} of $\hat{B}$. By $\llbracket \hat{B} \rrbracket_{\alpha} $ we denote the set of all \textnormal{$\alpha$-realizations} of $\hat{B}$. An output history $y \in \hat{B}(x)$ is called \textnormal{$\alpha$-realizable} for a fuzzy I/O behavior with input x, if there exists a realization $[\hat{b}]_{\alpha} \in \llbracket \hat{B} \rrbracket_{\alpha} $ with $y=\hat{b}(x)$.
\end{definition}

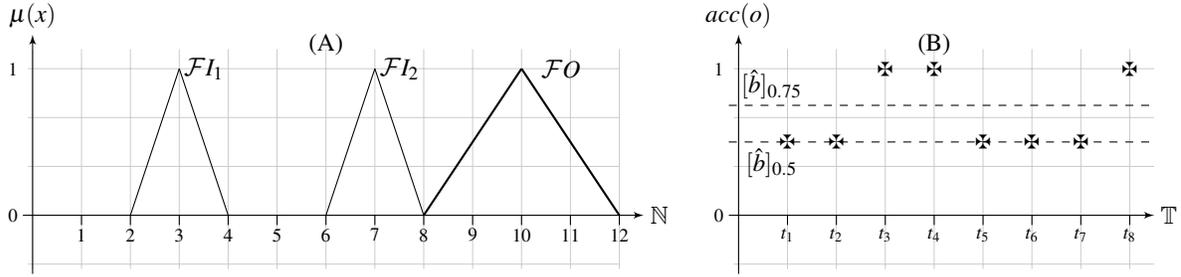
\begin{figure}[h!]
\centering
\vspace{0cm}
\begin{minipage}{.6\textwidth}
  \centering
  \begin{tikzpicture}[auto, node distance=0.5cm,>=latex',scale=0.65, transform shape]
  \colorlet{lightgray}{black!20}
  \draw[ultra thin,color=lightgray] (-0.1,-1.1) grid (12,3.4);
  \draw[->] (-0.2,0) -- (12.5,0) node[right] {\Large$\mathbb{N}$};
  \draw[->] (0,-1.2) -- (0,3.7) node[above] {\Large$\mu(x)$};
  \node at (-0.4,3){1};
  \node at (-0.4,0){0};
  \node at (1,-0.4){1};
  \draw (1,0) -- (1,-0.2) ;
  \node at (2,-0.4){2};
  \draw (2,0) -- (2,-0.2) ;
  \node at (3,-0.4){3};
  \draw (3,0) -- (3,-0.2) ;
  \node at (4,-0.4){4};
  \draw (4,0) -- (4,-0.2) ;
  \node at (5,-0.4){5};
  \draw (5,0) -- (5,-0.2) ;
  \node at (6,-0.4){6};
  \draw (6,0) -- (6,-0.2) ;
  \node at (7,-0.4){7};
  \draw (7,0) -- (7,-0.2) ;
  \node at (8,-0.4){8};
  \draw (8,0) -- (8,-0.2) ;
  \node at (9,-0.4){9};
  \draw (9,0) -- (9,-0.2) ;
  \node at (10,-0.4){10};
  \draw (10,0) -- (10,-0.2) ;
  \node at (11,-0.4){11};
  \draw (11,0) -- (11,-0.2) ;
  \node at (12,-0.4){12};
  \draw (12,0) -- (12,-0.2) ;
  \draw (2,0) -- (3,3);
  \draw (3,3) -- (4,0);
  \node at (3.5,3){\Large$\mathcal{F}I_1$};
  \draw (6,0) -- (7,3);  
  \draw (7,3) -- (8,0);
  \node at (7.5,3){\Large$\mathcal{F}I_2$};
  \draw[thick] (8,0) -- (10,3);  
  \draw[thick] (10,3) -- (12,0);
  \node at (10.8,3){\Large$\mathcal{F}O$};
  \node at (6,3.5){\Large (A)}; 	
  \end{tikzpicture}
  
\end{minipage}%
\begin{minipage}{.4\textwidth}
  \centering
  \begin{tikzpicture}[auto, node distance=0.5cm,>=latex',scale=0.65, transform shape]
  \colorlet{lightgray}{black!20}
  
  \draw[ultra thin,color=lightgray] (-0.1,-1.1) grid (8.5,3.4);
  \draw[->] (-0.2,0) -- (8.5,0) node[right] {\Large$\mathbb{T}$};
  \draw[->] (0,-1.2) -- (0,3.7) node[above] {\Large$acc(o)$};
  
  \node at (-0.4,3){1};
  \node at (-0.4,0){0};
  
  \node at (1,-0.4){$t_1$};
  \draw (1,0) -- (1,-0.2) ;
  \node at (2,-0.4){$t_2$};
  \draw (2,0) -- (2,-0.2) ;
  \node at (3,-0.4){$t_3$};
  \draw (3,0) -- (3,-0.2) ;
  \node at (4,-0.4){$t_4$};
  \draw (4,0) -- (4,-0.2) ;
  \node at (5,-0.4){$t_5$};
  \draw (5,0) -- (5,-0.2) ;
  \node at (6,-0.4){$t_6$};
  \draw (6,0) -- (6,-0.2) ;
  \node at (7,-0.4){$t_7$};
  \draw (7,0) -- (7,-0.2) ;
  \node at (8,-0.4){$t_8$};
  \draw (8,0) -- (8,-0.2) ;
   	
  \node at (1,1.5){$\maltese$}; 	
  \node at (2,1.5){$\maltese$}; 
  \node at (3,3){$\maltese$};
  \node at (4,3){$\maltese$};
  \node at (5,1.5){$\maltese$}; 	 	 		 	
  \node at (6,1.5){$\maltese$}; 	
  \node at (7,1.5){$\maltese$}; 	
  \node at (8,3){$\maltese$}; 	

  \node at (4,3.5){\Large (B)}; 	 	
   	
  \draw[dashed] (-0.2,2.25) -- node[near start, above, xshift=-13mm]{\Large $[\hat{b}]_{0.75}$} (8.5,2.25);
  \draw[dashed] (-0.2,1.5) -- node[near start, anchor=north, xshift=-13mm]{\Large $[\hat{b}]_{0.5}$} (8.5,1.5); 	 	
   
  \end{tikzpicture}
\end{minipage}
\caption{(A) Fuzzy type extension for a stateless adder. (B) $\alpha$-realization of stateful adder with fuzzy behavior $\hat{b}$}
\label{fig:fuzzyBehavior}
\vspace{0cm}
\end{figure}

\begin{example}[Stateful Fuzzy Behavior] Consider the following two programs (left: boolean, right:fuzzy) which is the stateful extension for the example \ref{ex:statelessBehavior}. 

\vspace{-0.3cm}
\begin{minipage}{.45\textwidth}
\begin{flalign*}
 \text{\textbf{fct} }   b=&(i_1: \text{I\textsubscript 1}, i_2: \text{I\textsubscript 2)} \mbox{ \textbf{out} } o:\text{O } \{ && \\
&  \langle \mbox{first}(i_1) + \mbox{ first}(i_2) \rangle \textcircled c &&\\
&  b(\mbox{rest}(i_1),\mbox{rest}(i_2)) \}  
\end{flalign*}
\end{minipage}%
\begin{minipage}{.5\textwidth}
\begin{align*}
\text{\textbf{fct} }  \hat{b}=&(i_1: \mathcal{F}I_1, i_2: \mathcal{F}I_2) \mbox{ \textbf{out} } o:\mathcal{F}O  \{ \\
& \langle \mbox{first}(i_1) + \mbox{ first}(i_2) \rangle \textcircled c \\
& b(\mbox{rest}(i_1),\mbox{rest}(i_2))\}  
\end{align*}
\end{minipage}
\vspace{0.2cm}

\noindent Now let $i_1=\langle 2,3,4,3,3,4,2,3\rangle$ be an input stream of fuzzy type $\mathcal{F}I_1$ and  $i_2=\langle 7,6,6,7,6,7,9,7\rangle$ another input stream of fuzzy type $\mathcal{F}I_2$. Then, the fuzzy behavior $\hat{b}(i_1,i_2)$ is $0.5$-realizable but it is not $0.75$-realizable as visualized in Figure \ref{fig:fuzzyBehavior}-B.  

Concluding this Section, we showed how to extend basic specification properties like realizability, in order to tackle with fuzzy behavior. In a similar way, theorem \ref{theorem:ExtensionPrinciple} and definition \ref{def:BehaviorExtension} provide the necessary tools for formalizing further specification properties such as safety, liveness, and fairness.     

\end{example}

\section{Related Work}
\label{sec:Related Work}

In the last decade many research efforts are recorded in literature \cite{Chechik2003,Cerny2014InterfaceSimulationDistances,NeubeckPhilipp2012,Henzinger2010SimulationDistances,Henzinger2014ModelMeasuringHybridSystems,Matthews2000FormalSpec,Matthews2002FuzzyConceptsAndFormalMethodesSample,Kwiatkowska2004}, where classical formal methods  have been extended with probabilistic, stochastic, distance measurement, and multi-valued logic techniques in order to deal with uncertainties in modeling component-based interactive systems. However, uncertainty has two distinct facets: randomness and fuzziness both of which play basic roles in human reasoning, decision making and concept formation \cite{Zadeh1977PossibilityVsProbability}. While the former handles partial knowledge (lack of essential information) the latter deals with partial truth (inability to characterize information). Thus, we intentionally leave probabilistic and stochastic systems outside the scope of this paper, concentrating instead on how to deal with partial truth. For the specification and development of interactive systems in consideration of probabilistic effects  we refer to Neubecks dissertation \cite{NeubeckPhilipp2012} where a theoretical framework for probabilistic systems is provided. 

Chechik et al. \cite{Chechik2003} introduces the concept of multi-valued model-checking and describes a multi-valued
symbolic model-checker, $\chi$Chek for analyzing models that contain uncertainty or inconsistency. They develop a modeling
language based on a generalization of Kripke structures, where both atomic propositions and transitions between states may take any of the truth values of a given multi-valued logic. In addition to the theoretical foundation they present a model-checking algorithm which is illustrated on some examples. Finally, the formalization of specification properties such us fairness in multi-valued model-checking is addressed. While Chechik et al. concentrate on logics with a finite set of truth values (a 3-valued logic is evaluated in their examples), we explore the case of continuous intervals of truth values. Furthermore, the concept of mapping strategies introduced in this paper enables the dynamic reconfiguration of specified intervals of truth values, which is also an extension to the aforementioned work.\looseness=-1             

With respect to formal specification based on fuzzy logic, Matthews et al. \cite{Matthews2000FormalSpec} suggests fuzzy set theory as a possible representation scheme to deal with uncertainty. The main contribution of their work is an extension of a set based specification language, namely Z. They develop a suitable fuzzy set notation within the existing syntax of Z. A summary of a toolkit is provided that defines the operators, measures and modifiers necessary for the manipulation of fuzzy sets and relations.  In further work \cite{Matthews2002FuzzyConceptsAndFormalMethodesSample}, Matthews illustrates how the toolkit can be used to specify a simple fuzzy expert system. However, their approach does not capture component interactions, which is the primary concern in this paper.\looseness=-1 

Cerny et al. \cite{Henzinger2010SimulationDistances} in a recent attempt pointed out that boolean notions of correctness are formalized by preorders on systems. To overcome the limitations of a two-valued logic, the authors introduce the notion of distances between two systems or between a system and a specification, and suggest quantitative simulation games as a framework for measuring such distances. They presented three particular distances: two for quantifying aspects of correct systems, namely coverage and robustness; and one for measuring the degree of correctness of an incorrect system. 
In a later work \cite{Cerny2014InterfaceSimulationDistances}, the same authors extend the quantitative notion of simulation distances to automata with inputs and outputs. The introduced interface distance, allows for measuring the desirability of an interface w.r.t. a given specification. In a direct comparison with the work presented in this paper one could say that both approaches pursue the same objective, namely to relax the boolean notion in formal specifications. However, the common objective is addressed by two distinct approaches. While Cerny et al. define for each property of interest a simulation distance and measure afterwards the deviation of all models, we rely on fuzzy set theory to soften the boolean notion. Hence, we suggest to formalize properties in terms of a-cuts and acceptance degrees on vague descriptions and measure to what degree a property of interest is fulfilled by concrete models (e.g. $a$-Realizabilty of two behaviors).\looseness=-1   

The restrictions of a two-valued logic are present also in systems with continuous behavior. Henzinger et al. presented in their recent paper \cite{Henzinger2014ModelMeasuringHybridSystems} a model measuring framework for the hybrid case, where distances are represented by parametrized hybrid automata. Actually, they address the same problem as described in \cite{Henzinger2010SimulationDistances} for the hybrid case. In our approach, we consciously decided for fuzzy set theory because of the fuzzification property which allows the generalization of a distinct theory to a continuous one. Thus, the introduced concepts in this paper can be easily generalized to continuous behaviors. An interesting future research objective would be to analyze the trade-off between fuzzy and hybrid approaches, in general. While hybrid automata make use of differential equations to describe a state,  fuzzy approaches use vague rules. What is the distance between fuzzy descriptions and differential equations?\looseness=-1 
\section{Conclusion}
\label{sec:Conclusion}
\begin{wrapfigure}{r}{8.3cm}
\vspace{-.5cm}
\includegraphics[scale=0.78]{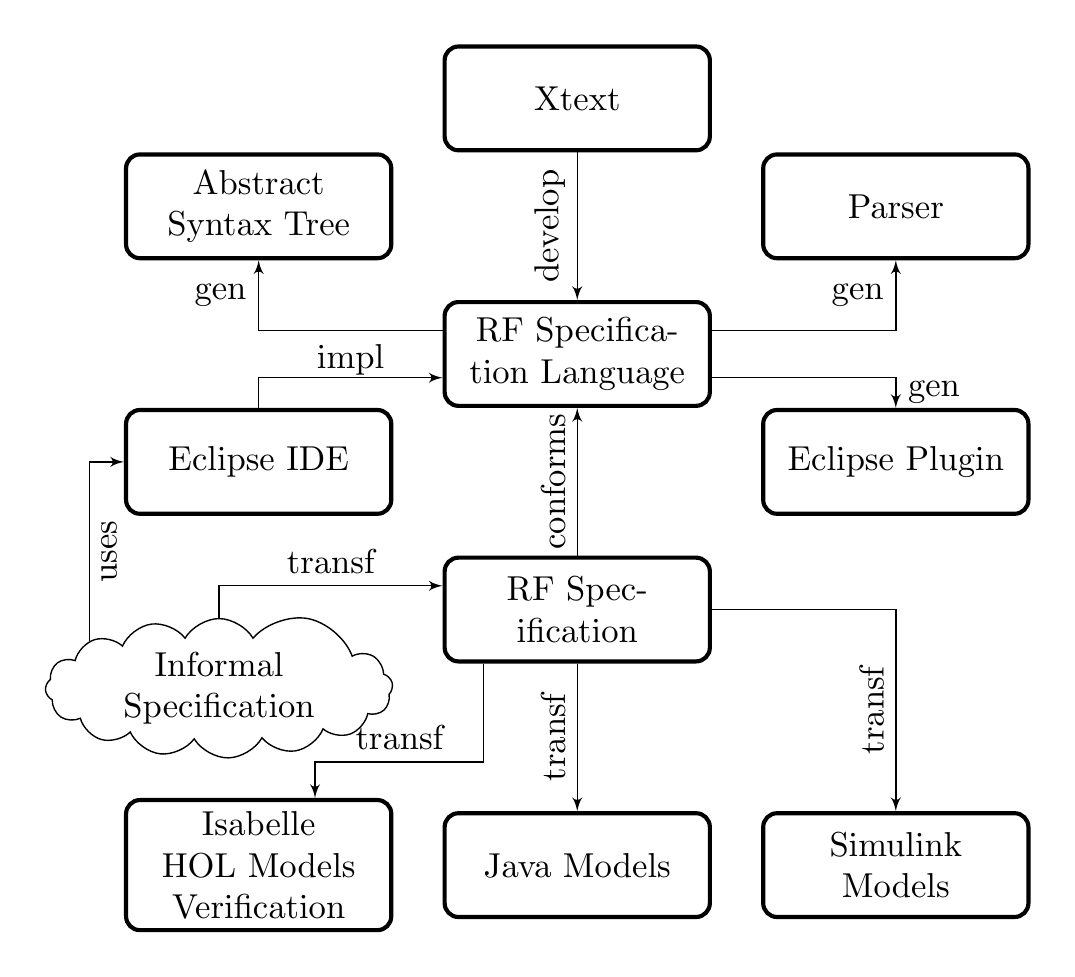}
\vspace{-0.9cm}
\caption{Tool Prototype}
\vspace{-0.5cm}
\label{fig:prototype}
\end{wrapfigure}

\paragraph{Tool Support.}
The intention of this work was not to present a concrete tool which is part of a tool demonstration but rather to establish the underlying theory required for the development of such a tool. Thus, we abstract away from the implementation details and present only an overview of a prototype under development depicted in Figure \ref{fig:prototype}. Xtext \cite{eysholdt2010xtext}, a framework for development of programming languages and DSLs is the starting point. It is used for the development of a model based specification language with support to the introduced concepts in chapters \ref{sec:FuzzyLogiconTopOfFocus},\ref{sec:Fuzzy Components}. The model based specification language generates the required parser and linker. Additionally, an eclipse plugin is generated which enables full support for the specification language inside the eclipse IDE. Hence, the integrated specification editor is used to transform the informal specification (requirements) into a formal specification which conforms to the developed language. Once, the informal requirements are formalized a series of model transformations becomes available. On the one hand, the specification can be transformed to executable models (Java and Simulink) which allows automated simulation for the system under development. On the other hand a generic theorem prover Isabelle \cite{Nipkow-Paulson-Wenzel:2002} is used for the verification and validation of system properties. Currently, there is only support for the introduced concepts in chapter \ref{sec:FuzzyLogiconTopOfFocus}, see for example Focus on Isabelle \cite{Spichkova08focuson}. In particular, support for formal verification of fuzzy component behavior is a major future research direction. Hence, a primary concern is to develop/adapt a fuzzy theory toolbox in Isabelle which enables fuzzy reasoning inside the framework.

\paragraph{Summary.}
In chapter \ref{sec:FuzzyLogiconTopOfFocus}, we introduced a specification technique based on fuzzy logic for interactive systems. In particular, we showed that a fuzzy rule based specification can be represented in terms of a black box view as a deterministic behavior and can be therefore modeled in a deterministic fashion by means of automata. The introduced technique is well suited for modeling especially user and environment interactions which are characterized by vagueness and uncertainty. The underlying Focus theory has been adapted to enable vague desriptions over fuzzy I/O ports. Finally, mapping strategy are introduced, which adapts fuzzy properties to the measured behavior over the I/O histories. Mapping strategies are well suited for formalizing self* properties.\looseness=-1      

In chapter \ref{sec:Fuzzy Components}, we introduced fuzzy components and fuzzy behavior of them. We established a basic notion for fuzzy types, channels and interfaces and provided basic operators on streams. A general method which enables the mapping of fuzzy input streams to fuzzy output streams over a set valued function is defined. The latter enables the modeling of fuzzy component behavior. Finally, we showed the fuzzy interpretation of basic specification properties like realizability.\looseness=-1 

\paragraph{Outlook.}
Concluding, we point out that our proposed method allows to capture certain system aspects which can not be represented by formal methods based on a two-valued logic. However, the work presented here is only an introduction towards a complete theory for fuzzy interactive systems. Basic system concepts as composition and decomposition, refinement, interface abstraction and architecture, to name only a few, have to be addressed in more detail. Last but not least from a more practical point of view specification techniques such as tables and diagrams and tool support in the form of AutoFocus \cite{AutoFocus} are future directions we have to go in order to set up more practical case studies to evaluate the expressiveness, completeness, and effectiveness of the introduced approach.\looseness=-1

\subsubsection*{Acknowledgments.}
The author address special thanks to Prof. Manfred Broy, Diego Marmsoler, Jonas Eckhardt, and Orestis Gkorgkas for their invaluable suggestions and the fruitful discussions on the topic.\looseness=-1

\bibliographystyle{eptcs}
\bibliography{paper}
\end{document}